\documentclass[runningheads,a4paper]{llncs}

\usepackage{amssymb}
\usepackage{graphicx}
\usepackage{color}

\usepackage{url}
\newcommand{\keywords}[1]{\par\addvspace\baselineskip
\noindent\keywordname\enspace\ignorespaces#1}

\begin{document}

\mainmatter  

\title{Dynamic Interaction Between Asset Prices and Bank Behavior: A Systemic Risk Perspective}

\titlerunning{Dynamic Interaction Between Asset Prices and Bank Behavior}

%
%
\author{
Aki-Hiro Sato$^1$ \and Paolo Tasca$^2$ \and Takashi Isogai$^3$
}
%
\authorrunning{
Aki-Hiro Sato and Paolo Tasca and Takashi Isogai
}

\institute{Graduate School of Informatics, Kyoto University, \\
Yoshida Honmachi, Sakyo-ku, Kyoto 606-8501 Japan 
\and
Research Centre, Central Office, Deutsche Bundesbank,\\
Wilhelm-Epstein-Stra\ss e 14 Frankfurt am Main 60431 Germany
\and
Financial System and Bank Examination Department, Bank of Japan \\
2-1-1 Nihonbashi-Hongokucho, Chuo-ku, Tokyo 103-0021, Japan \\
\url{http://ssuopt.amp.i.kyoto-u.ac.jp/akihiro/wiki_en/}
}

%
%

\toctitle{Dynamic Interaction Between Asset Prices and Bank Behavior}
\tocauthor{Aki-Hiro Sato, Paolo Tasca and Takashi Isogai}
\maketitle

\begin{abstract}
Systemic risk in banking systems remains a crucial issue that 
it has not been completely understood. In our toy model, banks are
exposed to two sources of risks, namely, market risk from their
investments in assets external to the banking system and credit risk from
their lending in the interbank market. By and large, both risks
increase during severe financial turmoil. Under this scenario, the
paper shows the conditions under which both the individual and
the systemic default tend to coincide. 
\keywords{banking system; capital adequacy ratio (CAR); procyclicality; agent-based model; financial market
}
\end{abstract}

\section{Introduction}
Understanding of the interplay among banks through several channels is
a crucial issue in the globalized world economy~\cite{Beale,Montagna}.
In general, banks obtain their profits from the difference between deposit
interest rates and interest rates in the interbank markets, stock
markets, credit markets, and so on. Of course, money flows and
interest rates are deeply interrelated to international economic conditions. 

It is recognized that systemic risks are created by
interconnection among banks. Helbing argues that systemic 
failures and extreme events are consequences of the highly 
interconnected systems and networked risks. He proposes 
a general theory to analyze, understand and manage systemic 
failures for various types of fields in 
socioeconomic-technological-environmental systems under 
a framework of Global Systems Science~\cite{Helbing}.
He further addresses a list of common drivers of systemic
instabilities that may destabilize anthropomorphic systems over
time. \textcolor{black}{These following drivers are to be considerably 
significant:} (1) increasing system sizes, (2) reduced redundancies
due to attempts to save resources (implying a loss of safety margins),
(3) denser networks (creating increasing interdependencies between
critical parts of the network), and (4) a high pace of innovation (producing
uncertainties or 'unknown unknowns').

These systemic risks probably \textcolor{black}{stem} from 
the positive feedback loop
among financial markets and banks' interactions. They can be formed by
several reasons such as leverage trading, trend follower's trading,
\textcolor{black}{loss} cut trading, bankruptcy of banks, 
bunched sales, housing market
and real economy, employment, excess concentration of wealth, a trade
imbalance, political power, extreme low interest rates, excessive or
lack of regulation and so on.

Several types of nonlinear positive feedback mechanisms can
concurrently trigger serious crashes damaging all the financial
systems. \textcolor{black}{This study collocates with that stream of
works that aim to} estimate systemic risks among
banks under exposures of risky assets traded in \textcolor{black}{the} 
stock market~\cite{Tasca2}.

Some existing studies \textcolor{black}{concern} 
risk propagation through lending and
borrowing networks. This approach focuses on interactions of debt and
credit among banks. 

For example, Iori et al. \textcolor{black}{analyze} the systemic risk in 
interbank money market by simulating the banks' lending
activities~\cite{Iori}. Their simulation model assumes that banks
carry liquidity risk caused by the maturity gap between funding and
investment activities. The model introduce\textcolor{black}{s} a feature
according to which banks pool this risk and further creates the
potential for one bank's crisis to propagate through the system.

\textcolor{black}{Furthermore,} the heterogeneity of banks
\textcolor{black}{is analyzed}, which can cause knock-on
effects in shock propagation, while the interbank market comprised of
homogeneous banks tends to stabilize absorbing
shocks. \textcolor{black}{This} model settings, two types of banks are
assumed: those with positive cash and those with negative
cash. Accordingly, they are classified as potential lenders and
potential borrowers. Banks invest their money first and lend the
remaining part as lending. The total demand in the market does not
always match the total supply. Hence, there exists default risk of
banks due to the shortage of liquidity, and shocks can be propagated
through the system.

As for the performance of the interbank market in its role as a safety
net, the insurance role of interbank lending prevails when banks are
homogeneous; higher reserve requirements can lead to a higher
incidence of bank failures. When banks are heterogeneous in average
liquidity or average size, contagion effects may arise. They found that
such effects can be of both a direct (i.e., knock-on from a failing
bank to its direct creditors) and an indirect (i.e., causing
criticality in the system as a whole) nature.  

Despite the potential to create contagion, they insist that inter-bank
lending always seems to stabilize the system: the elapsed time before
the first failure is \textcolor{black}{always} observed to increase with
connectivity. Their simulation results also indicate that
heterogeneity alone can contribute to instability.

Gai and Kapadia \textcolor{black}{study} that probability and potential impact of
contagion, which \textcolor{black}{are} influenced by aggregate and 
idiosyncratic shocks, change in network structure\textcolor{black}{,} 
and asset market liquidity~\cite{Gai}.

\textcolor{black}{A} model of contagion in arbitrary financial networks
\textcolor{black}{is developed by} using directed and weighted network
model to express the widespread transmission of shocks through
numerical simulation about shock transmission. 

The banks are linked together by their financial claims on
each other, including through interbank markets and payment systems.  
They model contagion stemming from unexpected shocks in complex
financial networks with arbitrary structure and then use numerical
simulations to illustrate and clarify the intuition underpinning our
analytic results. 

The result of simulation analysis suggests that financial systems
exhibit a robust-yet-fragile tendency: while the probability of
contagion may be low, the effects can be extremely widespread when
problems occur. Adverse aggregate shocks and liquidity risk amplify
the likelihood and extent of contagion.

\textcolor{black}{It is} also \textcolor{black}{clarified} 
why the resilience of the system in withstanding
fairly large shocks prior to 2007 should not have been taken as a
reliable guide to its future robustness. It means that we need more
flexible assumption when building network based model.  

The approach provides a first step towards modelling contagion risk
when true linkages are unknown. They \textcolor{black}{suggest a} 
further extension of the
analysis by relaxing the assumption that the defaulting bank is
randomly selected and, considering the implications of targeted
failure affecting big or highly connected interbank borrowers. As
mentioned, added realism could also be incorporated into the model by
using real balance sheets for each bank or endogenizing the formation
of the network. Such extension of the model would be beneficial from a
systemic risk research viewpoint. 

Haldane and May \textcolor{black}{study} possible effects of risk optimization by
individual financial institutions on the stability of the system as a
whole~\cite{Haldane}. They explore the interplay between complexity
and stability in deliberately simplified models of financial networks
to find some policy lessons with the explicit aim of minimizing
systemic risk. They \textcolor{black}{claim} that the network dynamics
of what might be called 'financial ecosystems' has parallels with
ecology\textcolor{black}{, where} too 
much complexity implies instability. The well-known arbitrage pricing
theories (APT) as well as other derivative pricing theories often
assume perfect competition, market liquidity, no-arbitrage and market
completeness. Crucially, these conditions are not always satisfied;
therefore, trading activities that assume these conditions can
destabilize markets, having possible \textcolor{black}{effect} on the dynamical
behavior, while such activities seem to be successful at an
initial stage. 

Haldane and May also delve into the shock propagation mechanism, applying 
network system approach originally developed in ecology. A financial system
is expressed as a network in which many banks are connected with
credit linkage, forming an inter-bank money market. An initial shock
that arose in some node are transmitted to other connected nodes when
their shock absorbing capacity of a node is insufficient to the
incoming shock. 

\textcolor{black}{In addition the} liquidity factor plays a major role
in the shock propagation. The losses in the value of bank external assets,
caused by a generalized fall in market prices, such liquidity shocks
amplify as more banks fail accordingly. Thus, relatively small initial
liquidity shocks have the potential to make strong contributions to
systemic risk. \textcolor{black}{Iori et al.} also emphasize the
cascading effect of shock 
propagation, in which diminished availability of interbank loans caused
serial failure of liquidity funding by banks. These complicated
interactions between nodes in a network cannot be clarified by the
traditional economic theory.

The traditional rationale for setting regulatory capital/liquidity
ratios is that idiosyncratic risks \textcolor{black}{are reduced} 
to the balance sheets of individual banks. Prudential regulation is
following in the footsteps of ecology, which has increasingly drawn on
a system-wide perspective when promoting and managing ecosystem
resilience.

\textcolor{black}{The scientists also} listed two policy implications
from their topological network analysis: the diversity across the
financial system and \textcolor{black}{the} modularity 
within the financial system. They warn that banks' balance sheets and
risk management systems became increasingly homogeneous; homogeneity
bred fragility. As for the modularity within the financial system,
\textcolor{black}{it} protects the systemic resilience of both natural and
constructed networks by limiting the potential for cascades. They
emphasized the importance of encouraging modularity and diversity in
banking ecosystems as a means of buttressing systemic resilience.

One of the authors (P.T.) proposed DebtRank to measure default risk of
banks. The paper assumes that bankruptcy may influence 
the balance sheet situation of other banks and that contagion effect may
happen~\cite{Tasca}. DebtRank is one of \textcolor{black}{the}
indicators to calculate the risk of the contagion effect for each
bank.

Reducing procyclicality and promoting \textcolor{black}{countercyclical} 
buffers \textcolor{black}{is} one
of \textcolor{black}{the most} important issues in the Basel III
Accords~\cite{BaselIII}. It provides a message that 
{\it ``One of the most destabilizing elements of the crisis has been
  the procyclical amplification of financial shocks throughout the
  banking system, financial markets and the broader economy.\textcolor{black}{``}}

In this study, we focus on correlations among financial assets. 
This may play a role of common factors and create
procyclicality. To do so, we study interactions between banking
systems and \textcolor{black}{the} financial market. Financial prices
fluctuate and show volatility clustering and volatility
synchronization. In fact, 
fluctuations of financial prices seem to be random, however,
volatilities of financial assets are sometimes synchronous. If many
banks have positions in financial assets, then their capital adequacy
ratios sometimes may vary synchronously. This is a common factor
effect of financial assets in balance sheet. In order to investigate
this scenario, we construct an agent-based model consisting
\textcolor{black}{of} banks and \textcolor{black}{the} financial market.

The linkage via underlying common factor is widely observed in the
financial market. Shocks prevail to the whole market in a short period
of time through \textcolor{black}{arbitrary} transactions by market
participants. Individual \textcolor{black}{assets return}, therefore
\textcolor{black}{they} tend to synchronize in terms of volatility
fluctuation: a large scale of volatility fluctuations are observed at
specific timing, resulting in a market shock event. 

Asset price fluctuation in the financial markets have been
studied by many researchers~\cite{Sornette,Bouchaud}. Stochastic
models as well as agent-based models are
well-studied~\cite{Micciche:02,Preis:06}. Some researchers pay a
significant attention to the network effects in the financial
markets~\cite{Bonanno:04,Vodenska}. From empirical studies on financial
time series, asset price fluctuations follow fat-tailed distributions.
This is often modeled as a random number drawn from a student $t$ 
distribution~\cite{Beale,Owen,Aki-Hiro:12}. However, since asset price
fluctuations are generated from trading by market participants,
high volatility regimes are not independent of banks behavior.

How does banks behavior affect financial markets and
\textcolor{black}{how} do the asset
price fluctuations influence banks' behavior? This forms a circular
causality. This is a main question of this study. To do so, we
consider a toy model of interaction among banks by means of an
agent-based model. We assume that banks have lending and
borrowing relationships with other banks and invest their money to
an asset. We focus on two viewpoints: capital adequacy
ratio and interaction between lending and borrowing network and
financial markets. 

One of the authors (T.I.) analyzed \textcolor{black}{the} Japanese stock
market by applying GARCH model to individual stock returns
separately~\cite{Isogai}. The result shows the market-wide
synchronization of extreme volatilities (larger than the 95th
percentile of the empirical distribution of individual volatilities),
which occurred mostly at crisis periods.

The purpose of this paper is to clarify influence of procyclicity. We
assume that financial assets in the balance sheet play a role as 
a common factor in a banking system. From an agent-based model 
consisting a banking system and financial market, we construct 
a model and measure correlation of capital adequacy ratio among banks. 
We found synchronous behavior of the capital adequacy ratio influenced 
by financial markets.

The rest of the paper is organized as follows\textcolor{black}{:}
Section \ref{sec:banks} defines a banking system\textcolor{black}{;}
Section \ref{sec:market-mechanism} defines behavior of market
participants\textcolor{black}{;} Section \ref{sec:debt-exposure} shows 
a model of chain-reaction bankruptcy\textcolor{black}{;} Section
\ref{sec:simulation} exhibits simulation results obtained from the
proposed agent-based model\textcolor{black}{;} Section
\ref{sec:discussion} tells some limitations of the 
model\textcolor{black}{;} Section \ref{sec:conclusion} is devoted to
\textcolor{black}{draw our} conclusions. 

\section{Banks}
\label{sec:banks}

\begin{figure}[!hbt]
\centering
\includegraphics[scale=0.5]{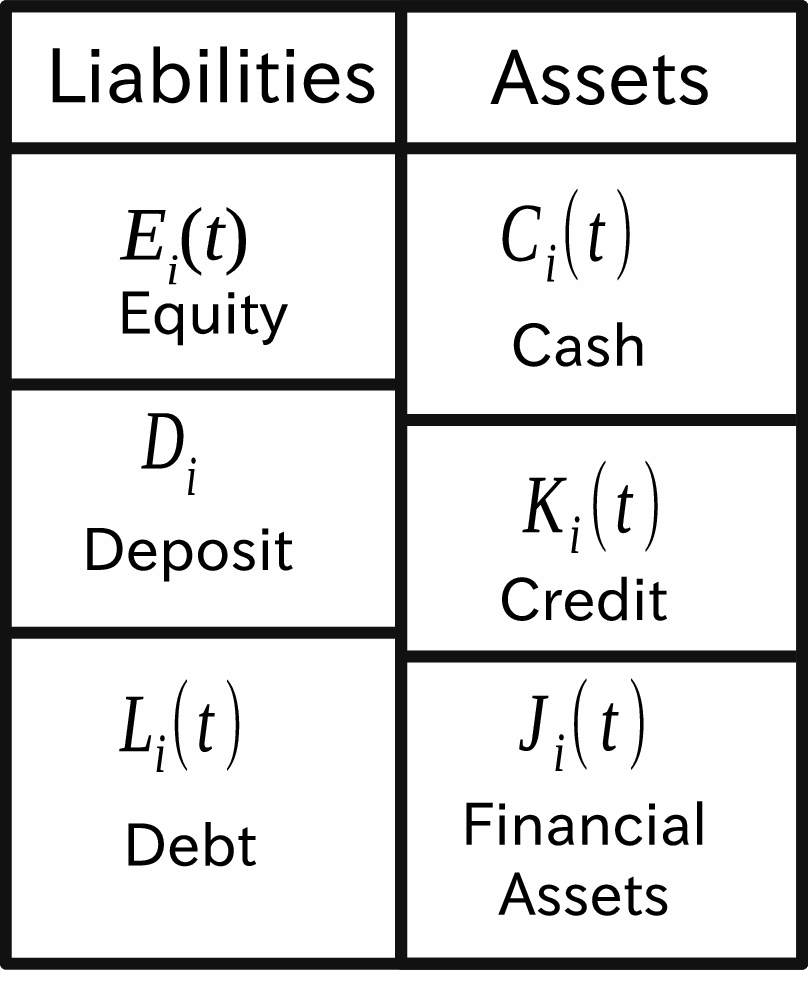}
\caption{A model of balance sheet. $E_i(t)$ represents Equity, $D_i(t)$ Deposit, $L_i(t)$ Debt, $C_i(t)$ cash, $K_i(t)$ Credit and $J_i(t)$ financial assets.}
\label{fig:balance_sheet}
\end{figure}

Let us consider \textcolor{black}{$N(t)$} banks which interplay one
another under lending-borrowing relationships and a financial market
\textcolor{black}{at time $t$}.

Suppose a balance sheet consisting of liabilities and assets 
in bank $i$ (See Fig. \ref{fig:balance_sheet}). Equity $E_i(t)$, Deposit
$D_i$, and Debt $L_i(t)$ are categorized as liabilities and Cash
$C_i(t)$, Credit $K_i(t)$ and Financial assets $J_i(t)$ as assets. 
We assume that the bank holds $n_i(t)$ units of risky assets with
their market price $S(t)$ and cash $C_i(t)$ at time $t$. Therefore, the
value of risky asset is estimated as $J_i(t) = n_i(t) S(t)$.

Since the bank is initially financed by their bank depositors, we
assume that $n_i(0) > 0$ and $C_i(0) > 0$. The bank deposit is
described as $D_i$, which is a constant value. If the $i$-th bank buys $V_i(t)$
units of risky asset then $n_i(t+\Delta t) = n_i(t) + V_i(t)$ and $C_i(t+\Delta t) = C_i(t) - V_i(t)S(t)$. If the $i$-th bank sells then
$n_i(t+\Delta t)=n_i(t) - V_i(t)$ and $C_i(t+\Delta t) = C_i(t) +
V_i(t)S(t)$. $V_i(t)$ represents volume traded by the bank $i$ at time $t$.

\textcolor{black}{Besides}, a lending and borrowing relationship exists
among banks. Such a 
relationship can be described as \textcolor{black}{an} asymmetric
weighted matrix. Let 
$W_{ij}(t)$ be expressed as a lending amount from the bank $i$
to the bank $j$ at time $t$. The debt of the $i$-th bank at time $t$ towards
other banks is estimated as $L_i(t) = \sum_{j=1}^N W_{ji}(t)$, and the 
bank credit of the $i$-th bank at time $t$ is estimated as $K_i(t) = \sum_{j=1}^N
W_{ij}(t)$. Therefore, the survival condition of the $i$-th bank is given by
\begin{equation}
C_i(t) + J_i(t) + K_i(t) > L_i(t) + D_i
\end{equation}
Namely, if $C_i(t) + J_i(t) + K_i(t) \leq L_i(t) + D_i$ then the
$i$-th bank goes bankruptcy. We assume that the counterparties of the 
$i$-th bank can receive $\rho \times 100$ percent ($0 < \rho < 1$) 
of their exposure. This can be expressed as temporal
development of interconnection among banks. 
The update rule of $W_{ji}(t)$ is given for all $i$ as
\begin{equation}
W_{\textcolor{black}{ji}}(t+\Delta t) =
\left\{
\begin{array}{ll}
W_{\textcolor{black}{ji}}(t) & (\mbox{If }C_i(t)+J_i(t)+K_i(t) \geq L_i(t)+D_i) \\
\rho W_{\textcolor{black}{ji}}\textcolor{black}{(t)} & (\mbox{Otherwise}) \\
\end{array}
\right. 
\label{eq:Wij-update}
\end{equation}

If we can simulate the agent-based model repeatedly, then from the
relative frequency we can estimate the default probability of the 
$i$-th bank as 
\begin{equation}
\mbox{Pr}[C_i(t) + J_i(t) \leq D_i + L_i(t) - K_i(t)] \approx 
\frac{M[C_i(t) + J_i(t) \leq D_i + L_i(t) - K_i(t)]}{M_{sim}},
\end{equation}
where $M_{sim}$ is the number of simulations, 
$M[C_i(t) + J_i(t) < D_i + L_i(t) - K_i(t)]$ is the number of
defaults for the $i$-th bank.

The bank $i$ must pay money to both depositors with deposit interest rates
and lenders with interest rates in the interbank market every
step. Such payments write 
\begin{equation}
C_i(t+\Delta t) = C_i(t) - \lambda_D D_i - \lambda_I L_i + \lambda_I K_i,
\end{equation}
where $\lambda_D$ represents the deposit interest rate, and $\lambda_I$
the interest rate in the interbank market. We assume that 
the interest rate is given by
\begin{equation}
\lambda = \Bigl(1+(\mbox{annual interest rate})\Bigr)^\frac{\Delta t}{365}-1,
\end{equation}
where $\Delta t$ is measured daily. In the case that 
the annual interest rate is 1\% and $\Delta t = 1$ [day], the daily
interest rate is estimated as $2.67262 \times 10^{-5}$.

Capital requirements are designed to ensure that banks hold enough
resources to absorb shocks to their balance sheets. A standard measure
of the health of individual banks is their capital adequacy ratio (CAR). 
Introduced in 1988 with the Basel I Accords, the CAR is calculated as
the total regulatory capital of a bank divided by its risk-weighted
assets. The Basel II revision refined the calculation of risk weights
and incorporated three major components of risk: credit, operational,
and market risk. The Basel II revision also set the minimum CAR at 8
percent for international banks and at 4 percent for domestic banks. 
Conservatively-run banks tend to have high CARs to cushion
against higher losses. In addition, Basel III revision introduced that
a Total Capital Ratio to total Risk Weighted Assets should be 
larger than 8 \% and that the common Equity Tier 1 to risky asset ratio is
greater than 4.5\%.

Bank capital to assets is the ratio of bank capital and reserves to
total assets. Capital and reserves include funds contributed by
owners, retained earnings, general and special reserves, provisions,
and valuation adjustments. Capital includes tier 1 capital (paid-up
shares and common stock), which is a common feature in 
\textcolor{black}{the} banking systems \textcolor{black}{all over the
  world} and total regulatory capital, 
which includes several 
specified types of subordinated debt instruments that need not
\textcolor{black}{to} be
repaid if the funds are required to maintain minimum capital levels
(these comprise tier 2 and tier 3 capital). Total assets include all
non-financial and financial assets. CAR is defined as
\begin{equation}
CAR = \frac{(\mbox{Tier 1 capital})+(\mbox{Tier 2 capital})}{(\mbox{Risk weighted assets})},
\end{equation}
where Tier 1 capital $T_1$ is defined as 
\begin{eqnarray}
\nonumber
(\mbox{Tier 1 capital}) &=& \mbox{(paid up capital)}+\mbox{(statutory reserves)} \\
\nonumber
&+&\mbox{(disclosed free reserves)}\\
\nonumber
&-&\mbox{(equity investment in subsidiary)}\\
\nonumber
&-&\mbox{(intangible assets)} \\
\nonumber
&-&\mbox{(current and b/f losses)},
\end{eqnarray}
and Tier 2 capital $T_2$ as 
\begin{eqnarray}
\nonumber
\mbox{(Tier 2 capital)} &=& \mbox{(Undisclosed Reserves)}\\
\nonumber
&+&\mbox{(General Loss reserves)}\\
\nonumber
&+&\mbox{(hybrid debt capital and subordinated debts)}.
\end{eqnarray}
The risk-weight depends on kinds of assets. In the case of cash and
government securities, the weight is 0\%. Mortgage loans have 50\%
(Basel I) weight or 35\% (Basel II). Other loans and assets have 100\% 
weight (Basel I) or 75\% to 150\% (Basel II). Stocks have 100\% weight
(Basel I and II). 

In the case of our model, we assume that the total capital adequacy ratio is 
approximated as 
\begin{equation}
CAR_i(t) = \frac{C_i(t)+J_i(t)+K_i(t)-L_i(t)-D_i}{J_i(t)+K_i(t)}
\times 100 (\%),
\end{equation}
and that the Tier1 common equity adequacy ratio is approximated by a ratio of
Tier1 common equity (Cash and Common Stocks) to risk weighted assets and operational risk;
\begin{equation}
CEAR_i(t) = \frac{C_i(t)}{J_i(t)+(1+\lambda_I c)K_i(t) + c|y_i(t)|V_i(t)S(t)}
\times 100 (\%),
\end{equation}
where $c$ is a positive constant less than 1. The Basel II Accords
\textcolor{black}{forced} the requirement such that $CAR_i(t) \geq 8$ (\%). 
The Basel III Accords have required $CEAR_i(t) \geq 4.5$ (\%) 
since 3Q in 2013 in addition to this. 

This is a model of a banking system. The variables in this model are 
listed in Tab. \ref{tab:variables}.

\begin{table}[!hbt]
\centering
\caption{Variables of banks and a market.}
\label{tab:variables}
\begin{tabular}{lll}
\hline
\multicolumn{1}{c}{variables} & \multicolumn{1}{c}{items} \\
\hline
$t$ & time \\
\textcolor{black}{$N(t)$} & \textcolor{black}{The number of market participants} \\
\textcolor{black}{$N_{tf}(t)$} & \textcolor{black}{The number of trend followers} \\
$S(t)$ & Market price of risky assets \\
$E_i(t)$ & Equity \\
$D_i$ & Deposit \\
$L_i(t)$ & Debt \\
$C_i(t)$ & Cash \\
$K_i(t)$ & Credit \\
$J_i(t)$ & Financial assets \\
$W_{ij}(t)$ & Lending amount from bank $i$ to bank $j$ \\
$n_i(t)$ & Holding volume of financial assets \\
$V_i(t)$ & Traded volumes \\
$y_i(t)$ & investment attitude \\
\hline
\end{tabular}
\end{table}

\begin{figure}[!hbt]
\centering
\includegraphics[scale=0.74]{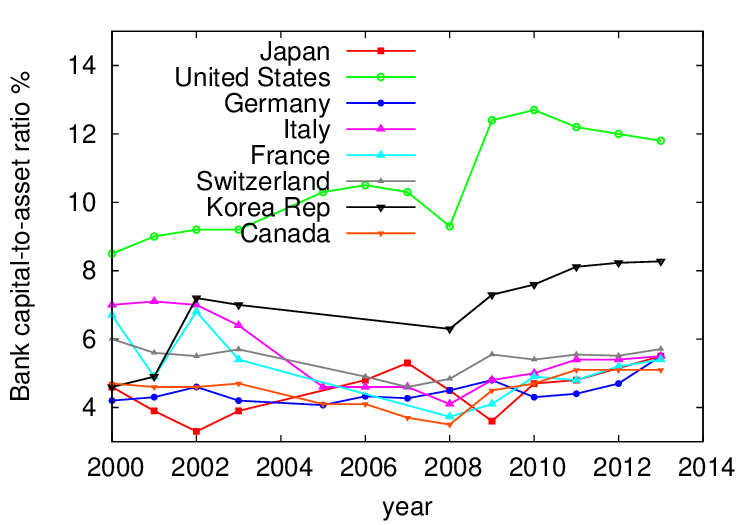}
\caption{COLOR ONLINE.
Averaged bank capital adequacy ratio in 8 countries (Japan,
  United States, Germany, Italy, France, Switzerland, Korea Rep and
  Canada) for a period from 2000 to 2013.}
\label{fig:CAR}
\end{figure}

Fig. \ref{fig:CAR} shows averaged CAR in 8 countries from year 2000,
\textcolor{black}{where} data can be downloaded from Worldbank's
DataBank~\cite{databank}.  
The graph tells us that the averaged CAR varies in time and that
\textcolor{black}{it fluctuates} in a range from 4.0 to 12.0. The
average CAR of United 
States \textcolor{black}{maintains} more than 8 \%. France, Germany and
Japan are less than 6 \% from 2000 to 2010. Canada and South Korea
take more than 6 \%.

However, these values are averages over the country. A specific bank
takes larger values than the average. For example, although the
Japanese averages are less than 5 \%, Japanese three mega banks
(Mizuho Financial Group, Mitsubishi Tokyo Financial Group and Sumitomo
Mitsui Financial Group) take larger values; Mizuho Financial Group shows 
15.09\% total capital ratio as of December 31, 2014~\cite{MizuhoFG:2014} 
(See Tab. \ref{tab:MizuhoFG}). Mitsubishi Tokyo Financial Group also 
shows 15.39\% total capital ratio as of September 31, 2014~\cite{MTFG:2014}
(See Tab. \ref{tab:MTFG}). Sumitomo Mitsui Financial
Group shows 16.79\% total capital ratio as of December 2014~\cite{SMFG:2014}
(See Tab. \ref{tab:SMFG}). The current capital-to-asset ratios reported by 
Worldbank are averaged over banks belonging to countries 
not regarding total amounts of banks equity and assets. It may be
necessary to compare the capital adequacy ratios in accordance with sizes
of banks' capital and risk weighted assets.

\begin{table}[!hbt]
\caption{Mizuho Financial Group, as of December 31, 2014.}
\label{tab:MizuhoFG}
\begin{tabular}{|l|l|l|}
\hline
\multicolumn{1}{|c|}{Items} & \multicolumn{1}{|c|}{Amount} & Basel III Template No. \\
\hline
Tier1 capital & 7,481,242M JPY & 45 \\
\hline
Teir2 capital & 2,181,862M JPY & 58 \\
\hline
Total capital & 9,663,105M JPY & 59 \\
\hline
Risk weighted asset & 64,023,907M JPY & 60 \\
\hline
\hline
Tier1 CAR & 11.68\% & 62 \\
\hline
Tier1 CAR (Common Equity Teir1) & 9.25\% & 61 \\
\hline
Total CAR & 15.09\% & 63 \\
\hline
\end{tabular}
\\

\caption{Mitsubishi Tokyo Financial Group, as of September 31, 2014.}
\label{tab:MTFG}
\begin{tabular}{|l|l|l|}
\hline
\multicolumn{1}{|c|}{Items} & \multicolumn{1}{|c|}{Amount} & Basel III Template No. \\
\hline
Tier1 capital & 12,726,118M JPY & 45 \\
\hline
Tier2 capital & 3,313,073M JPY & 58 \\
\hline
Total capital & 16,039,191M JPY & 59 \\
\hline
Risk weighted asset & 104,160,164M JPY & 60 \\
\hline
\hline
Tier1 CAR & 12.21\% & 62 \\
\hline
Tier1 CAR (common Equity Tier 1) & 10.97\% & 61 \\
\hline
Total CAR & 15.39\%  & 63 \\
\hline
\end{tabular}
\\

\caption{Sumitomo Mitsui Financial Group, as of December 31, 2014.}
\label{tab:SMFG}
\begin{tabular}{|l|l|l|}
\hline
\multicolumn{1}{|c|}{Items} & \multicolumn{1}{|c|}{Amount} & Basel III Template No. \\
\hline
Tier1 capital & 8,366,228M JPY & 45 \\
\hline
Tier2 capital & 2,547,949M JPY & 58 \\
\hline
Total capital & 10,914,178M JPY & 59 \\
\hline
Risk weighted asset & 64,992,642M JPY & 60 \\
\hline
\hline
Tier1 CAR & 12.87\% & 62 \\
\hline
Tier1 CAR (common Equity Tier 1) & 11.17\% & 61 \\
\hline
Total CAR & 16.79\%  & 63 \\
\hline
\end{tabular}
\end{table}

\section{Market mechanism}
\label{sec:market-mechanism}
The risky assets are traded through a common market. The bank traders
buy and sell their risky assets. For the sake of simplicity, we assume
that the investment attitude in the financial market is determined on the
basis of the last change in the market price. 

The market participants are classified into two types; trend followers
and contrarians. The trend followers are traders who want to buy (sell) assets
when their price goes up (down). The contrarians are traders who want to buy 
(sell) assets when their price goes down (up). \textcolor{black}{Suppose
that $N(t)$ banks trade a single asset at time $t$}. It is assumed
that the banks can take three investment attitudes coded as three
states (buying: 1, selling: -1, and waiting: 0):
\begin{equation}
y_i(t+\Delta t) = 
\left\{
\begin{array}{lcl}
1 & \mbox{with probability} & p_{i}^{(+)}(R(t)) \\
0 & \mbox{with probability} & 1 - p_{i}^{(+)}(R(t)) - p_{i}^{(-)}(R(t)) \\
-1 & \mbox{with probability} & p_{i}^{(-)}(R(t)) \\
\end{array}
\right.,
\end{equation}
where
\begin{eqnarray}
p^{(+)}(R) &=& \frac{1}{2}\mbox{erfc}\Bigl(\frac{\theta_{2i} - a_i R}{\sqrt{2}\sigma}\Bigr), \\
p^{(-)}(R) &=& \frac{1}{2}\mbox{erfc}\Bigl(\frac{a_i R-\theta_{1i}}{\sqrt{2}\sigma}\Bigr).
\end{eqnarray}
$a_i$ is a parameter which determines behavior of banks. If $a_i > 0$, then the $i$-th bank is a trend follower. If $a_i < 0$, then the $i$-th bank is a 
contrarian. $\theta_{1i}$ and $\theta_{2i}$ are parameters of the
$i$-th bank ($\theta_{1i} < \theta_{2i}$). The value of $\sigma (> 0)$
represents the uncertainty of decision. 

\textcolor{black}{In fact, there are some categories of banks such as 
investment, retail and central banks. Central banks
sometimes trade stocks for non-profit making but policy-oriented 
purposes (e.g., asset purchase for monetary easing). 
Investment banks and retail banks have different risk preference to financial
markets. These differences can be tuned by parameters $a_i$,
$\theta_{1i}$ and $\theta_{2i}$. The parameters strongly depend on
market conditions and banks' risk appetite. However, we do not have
enough information about \textcolor{black}{banks'} risk appetite in
general. We need to infer model parameters from available information
about macro economic conditions, bank balance sheet, and market sentiments. 
This problem arises other problems that are not meant to be
investigated in this paper. Parameter estimations should be seen as
an important future tasks.} 

\begin{figure}[!hbt]
\centering
\includegraphics[scale=0.67]{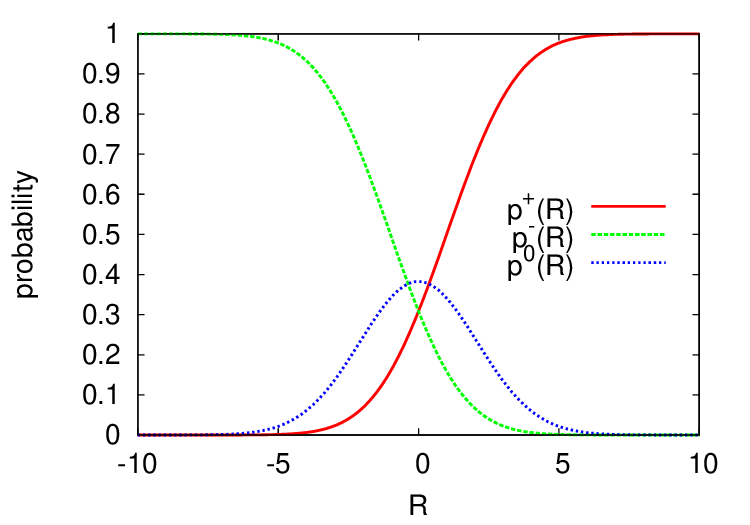}(a)
\includegraphics[scale=0.67]{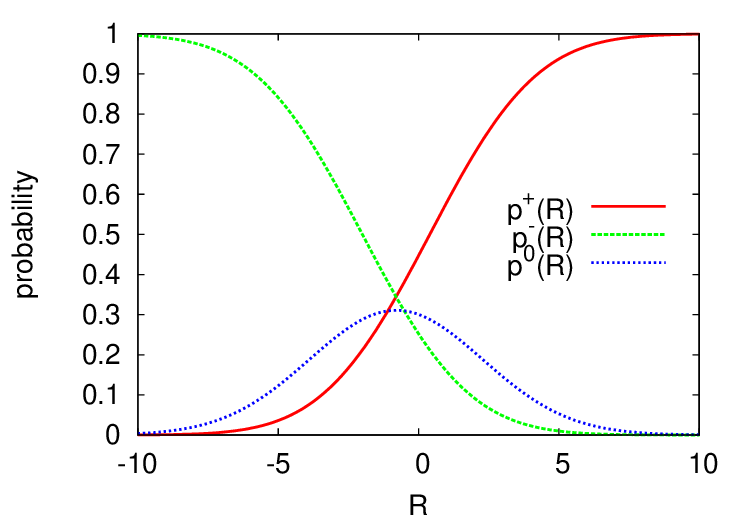}(b)
\caption{COLOR ONLINE. 
The relationship between a price change and probabilities of
  investment attitudes at (a) $\theta_{1i}=-1.0$, $\theta_{2i}=1.0$,
  $a_i=1.0$, and $\sigma=2.0$. (b) at $\theta_{1i}=-2.0$, $\theta_{2i}=0.4$,
  $a_i=1.0$, and $\sigma=3.0$.}
\label{fig:prob}
\end{figure}

These are models of response curves between perception (price change)
and three types of investment attitudes. We use these curves that
relate a price change to investment attitude. Fig. \ref{fig:prob}
shows the probabilities for three investment attitude. The
probabilities can be adjusted by using $\theta_{1i}$, $\theta_{2i}$,
$a$ and $\sigma$. Specifically, $\theta_{1i}$ and $\theta_{2i}$ are
parameters to describe a range of unresponsiveness to the price
change. While $\theta_{1i}$ normally takes negative values, $\theta_{2i}$ takes
positive values. The mode of $p^{(0)}(R)$ is equivalent to
$(\theta_{1i}+\theta_{2i})/2$. If we differ $|\theta_1|$ from
$|\theta_2|$, then we can express asymmetric response of price changes
to investment attitudes. This is also understood from probit-logit
reasoning.

Furthermore, it is assumed that excess demand for a risky asset
$\sum_{i=1}^N V_i(t)y_i(t)$ drives the market price. The volume traded
by the bank $i$ is assumed to be proportional to its amount of equity:
\begin{equation}
V_i(t) = \eta \Bigl(C_i(t)+J_i(t)+K_i(t)-L_i(t)-D_i\Bigr),
\end{equation}
where $\eta$ is an investment ratio taking from 0 to 1. 
\textcolor{black}{There is a general strategic framework to control
risk from price fluctuations in banking based on an amount of bank's
risk-based capital. In general, banks should trade risky assets
under the limit of its economic capital. This is a fundamental requirement 
in economic capital management~\cite{Elizalde:2007}. The risk-based capital
is basically linked with the amount of equity, although it is complicated 
to calculate risk-based capital exactly from balance sheet data. Therefore, 
we use an amount of equity to determine the volume traded by the bank.}

To guarantee positive market prices, the following log return is chosen:
\begin{equation}
R(t) = \log S(t+\Delta t) - \log S(t),
\label{eq:log-ret}
\end{equation}
and the log returns are proportional to the excess demand,
\begin{equation}
R(t) = \gamma \sum_{i=1}^N V_i(t)y_i(t),
\label{eq:log-ret2}
\end{equation}
where $\gamma$ is a positive constant to represent the response of the
return to the excess demand. This constant is associated with
price elasticity. 

From Eqs. (\ref{eq:log-ret}) and (\ref{eq:log-ret2}), we have
\begin{equation}
S(t+\Delta t) = S(t) \exp\Bigl(\gamma \sum_{i=1}^N V_i(t)y_i(t) \Bigr).
\end{equation}

After the bank trades stocks, the bank's amount of cash and the holding
number of stocks is updated. For the \textcolor{black}{buyer} side, if
$C_i(t) \geq V_i(t)S(t)$ then
\begin{equation}
C_i(t+\Delta t) = C_i(t) - V_i(t)S(t), \quad n_i(t+\Delta t) = n_i(t)
+ V_i(t).
\end{equation}
For the \textcolor{black}{seller} side, if $n_i(t) \geq V_i(t)$ then
\begin{equation}
C_i(t+\Delta t) = C_i(t) + V_i(t)S(t), \quad n_i(t+\Delta t) = n_i(t)
- V_i(t).
\end{equation}

\section{Debt exposure}
\label{sec:debt-exposure}
The $i$-th bank has equity $E_i(t)=C_i(t)+J_i(t)+K_i(t)-L_i(t)-D_i(t)$
at time $t$. If $E_i(t)$ is less than zero, we define that the $i$-th
bank goes bankrupt. Thus, it would be useful to formalize the Default Event
(DE) such that
\begin{equation}
DE := E = 0 := CAR(J+K)/100\%,
\end{equation}
which can be also seen as a function in terms of $CAR$.

Let us assume that a symmetric weighted matrix
$W_{ij}(t)$ describes a lending and borrowing relationship among banks
at time $t$. The debt of the \textcolor{black}{$j$-th} bank is estimated
as $L_j(t) = \sum_{i=1}^N W_{ij}(t)$. The default impact of the bank 
\textcolor{black}{$i$} at time $t$ is denoted as $W_{ij}(t)$. 
\textcolor{black}{In order to estimate the worst case scenario, 
we assume that $\rho$ in Eq. (\ref{eq:Wij-update}) 
is set as zero ($\rho=0$). If we use a nonzero value for $\rho$, then
we can simulate the case where some percentage of the debt of $j$-th
bank to $i$-th bank can be collectible.}

Let $Q_i^{(n)}$ be a cumulative loss of the bank $i$ in the $n$-th iteration.
If we introduce a binary variable $h_j^{(n)}$ such that $h_j^{(n)}=0$
(normal) and $h_j^{(n)}=1$ (default), then the cumulative losses of
the bank $i$ can be calculated as
\begin{equation}
Q_i^{(n+1)} = Q_i^{(n)} + \sum_{j=1}^N W_{ij}^{(n)}h_j^{(n)},
\end{equation}
where $W_{ij}^{(n)}$ is updated as
\begin{equation}
W_{ij}^{(n)} =
\left\{
\begin{array}{ll}
W_{ij}^{(n-1)} & (h_j^{(n-1)} = 0) \\
\textcolor{black}{\rho W_{ij}^{(n-1)}} & (h_j^{(n-1)} = 1)
\end{array}
\right..
\end{equation}
with the initial conditions given by
\begin{equation}
Q_i^{(0)} = 0, \quad W_{ij}^{(0)} = W_{ij}(t).
\end{equation}
If the cumulative loss $Q_i^{(n)}$ of the bank $i$ becomes greater than
its equity $E_i(t)$ 
\begin{equation}
Q_i^{(n)} > E_i(t),
\end{equation}
then we recognize that the bank $i$ goes bankrupt and set
$h_i^{(n+1)}=1$ (it becomes default) otherwise
$h_i^{(n+1)}=h_i^{(n)}$. We obtain the new matrix at time $t+\Delta t$
as $W_{ij}(t+\Delta t) = W_{ij}^{(\infty)}$.

Our interest is to understand the total losses over all the banks, which
is denoted as $H(t)$, triggered by the default of the bank $j$, 
which is estimated as the total losses
\begin{eqnarray}
H_i(t)&=&\sum_{j\neq i}h_j^{(\infty)}\Bigl(C_j(t)+J_j(t)+K_j(t)\Bigr), \\
H(t)&=&\sum_{i=1}^{N}H_i(t),
\end{eqnarray}
under the initial condition $h_i^{(0)}=0 \quad (i\neq j)$. $C_j(t)+J_j(t)+K_j(t)$
is assumed to be the economic value of the bank $j$.

\section{Simulation}
\label{sec:simulation}
In order to compute this algorithm, we set parameters as shown in
Tab. \ref{tab:parameters}. \textcolor{black}{We sample parameters from
a uniform distribution. In fact, some relationships between
parameters do exist and they are time-dependent. However, the
purpose of this numerical simulation is to show the interplay
between financial markets and the trend followers and how it can
play a role of the positive feedback mechanism and procyclical motion of
market prices. If we have detailed data on banks, we can set more
realistic parameters. Moreover, it is difficult to obtain
information about the behavior of all the banks, as
mentioned above. We should consider a way to set reliable parameters
from available information about banks balance sheets, however, such
an extension makes a problem more complicated. The is an important
issue to be analyzed and discussed in future studies.}

\textcolor{black}{In our numerical simulation, we assume that the
interest rate (5\%) is higher than the deposit interest rate (1\%)
but one is not much higher than the other. The current parameter model
setting is the toughest case scenario. If we set interest
rates higher than the deposit rates, then we can simulate a safer case
scenario than the assumptions given by the model presented.}

\begin{table}[!hbt]
\caption{Model parameters}
\label{tab:parameters}
\centering
\begin{tabular}{ll}
\hline
The initial number of banks & \textcolor{black}{$N(0)=100$} \\
\textcolor{black}{The initial market price} & \textcolor{black}{$S(0)=1$} \\
$\theta_{1i}$ & sampled from the uniform distribution $U(-1.0,-0.3)$ \\ 
$\theta_{2i}$ & sampled from the uniform distribution $U(0.3,1.0)$ \\ 
$\bar{a}_i$ & sampled from the uniform distribution $U(a_0,a_0+2.1)$ \\
\textcolor{black}{$a_i(t)$} & \textcolor{black}{$\bar{a}_i + a'_i(t)$, where
$a'_i(t)$ is sampled from $U(-\sigma_a, \sigma_a)$} \\
$\sigma$ & sampled from the uniform distribution $U(3.0,4.0)$ \\
$C_i(0)$ & sampled from the uniform distribution $U(2000,3000)$ \\
$n_i(0)$ & sampled from the uniform distribution $U(2000,3000)$ \\
$W_{ij}(0)$ & sampled from the uniform distribution $U(100,500)$ \\
$D_i$ & sampled from the uniform distribution $U(100,200)$ \\
Deposit interest rate & $\lambda_D=2.7262\times 10^{-5} (1\%)$ \\
Interbank interest rate & $\lambda_I=1.3368\times 10^{-4} (5\%)$ \\
Investment ratio to equity & $\eta=0.001$ \\
Price elasticity constant & $\gamma=0.1$ \\
\textcolor{black}{Randomness of $a_i(t)$} & \textcolor{black}{$\sigma_a=0.3$} \\
recovery ratio & $\rho = 0.0$ \\
Averaged number of links & 6 \\
Lending money $W_{ij}$ & sampled from the uniform distribution $U(100,500)$ \\
\hline
\end{tabular}
\end{table}

We also selected the total cash possessed by banks at time $t$ defined
as
\begin{equation}
C_{total} = \sum_{i=1}^N C_i(t),
\end{equation}
and the total number of stocks held by banks at time $t$ computed
by 
\begin{equation}
n_{total} = \sum_{i=1}^N n_i(t),
\end{equation}
as representative quantities describing conditions of
banks. The ratio of the trend followers to the total traders
\begin{equation}
\textcolor{black}{\alpha(t) = \frac{N_{tf}(t)}{N(t)}},
\end{equation}
where \textcolor{black}{$N_{tf}(t)$} is the number of trend
followers. \textcolor{black}{Fraction $\alpha(t)$} is used  
as a parameter to characterize the financial market 
\textcolor{black}{at time $t$. We can control $\alpha(t)$ by changing
$a_i(t)$ for every bank $i$ at time $t$. Usually, the fraction of the
number of trend followers to the total number of market participants
varies in time due to the evolution of the mechanism of participants' trading
strategies. To consider the temporal dependence of fraction $\alpha(t)$,
we introduced some randomness to $a_i(t) = \bar{a}_i + a_i'(t)$, where
$a'_i(t)$ is sampled from the uniform distribution $U(-\sigma_a,
\sigma_a)$. Here $\sigma_a$ is a positive
constant.}

\begin{figure}[!hbt]
\centering
\includegraphics[scale=0.43]{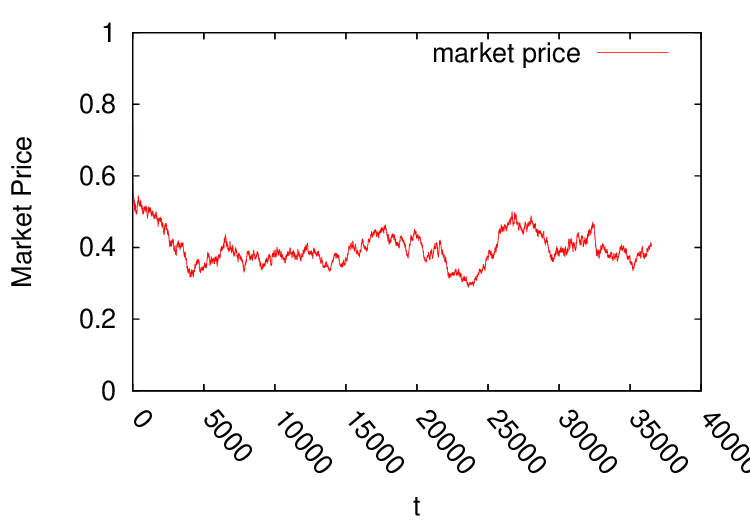}(a)
\includegraphics[scale=0.43]{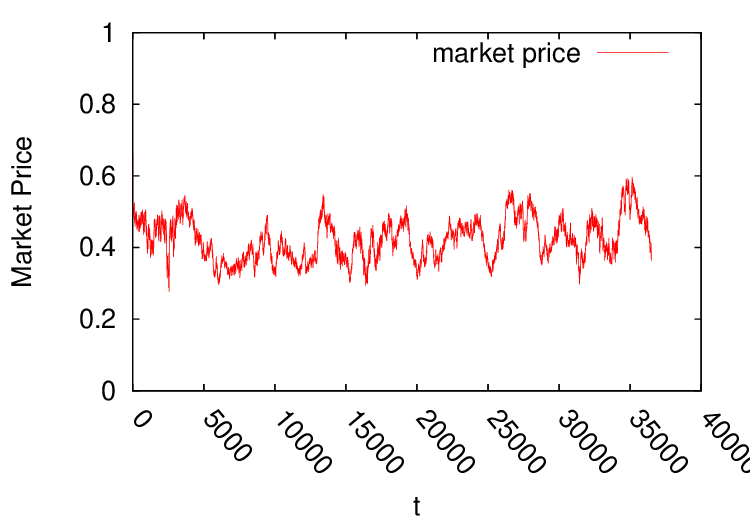}(b)
\includegraphics[scale=0.43]{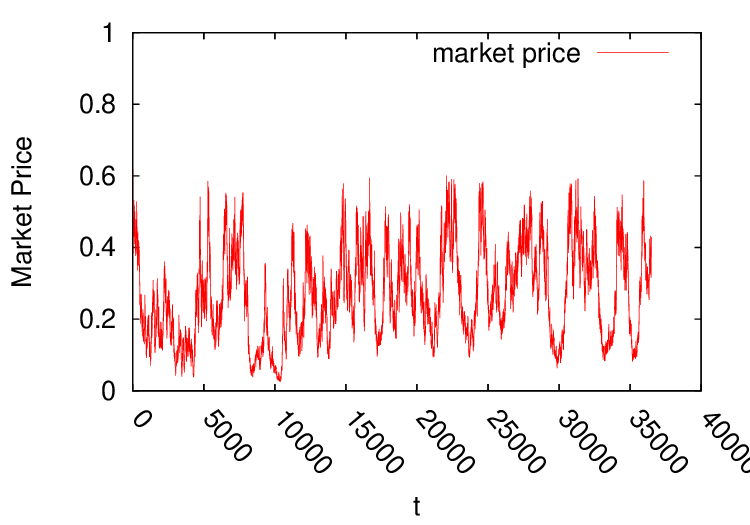}(c)
\includegraphics[scale=0.43]{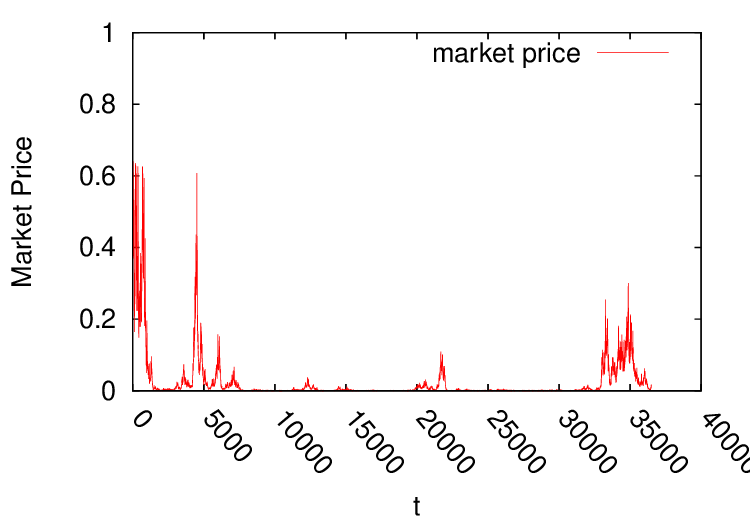}(d)
\caption{COLOR ONLINE. 
(a) The market price at \textcolor{black}{(a) contrarians-dominant market, 
(b) contrarians-predominant-market, (c) trend-followers-predominant market, 
and (d) trend-followers-dominant market}.}
\label{fig:markets}
\end{figure}

\begin{figure}[!hbt]
\centering
\includegraphics[scale=0.43]{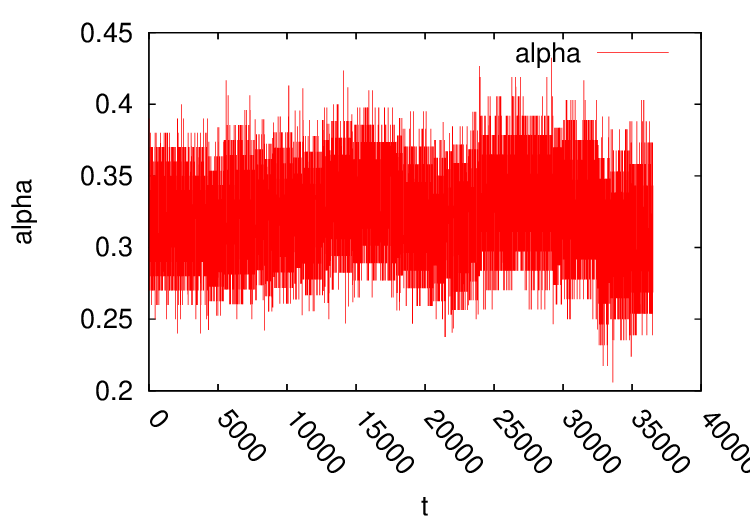}(a)
\includegraphics[scale=0.43]{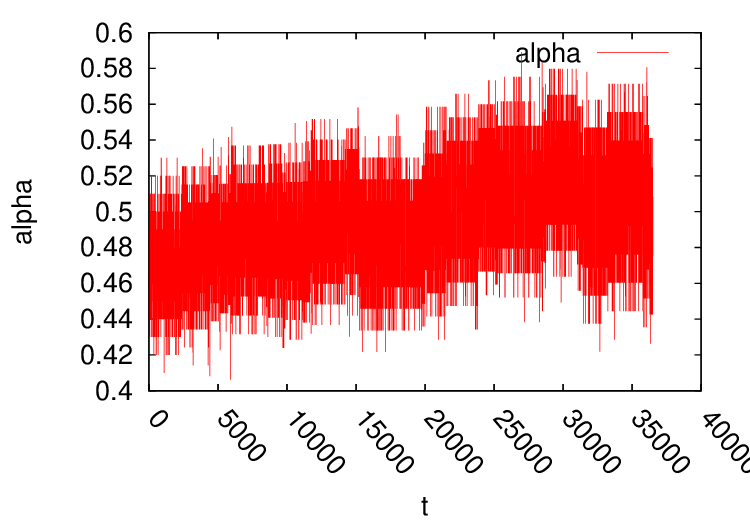}(b)
\includegraphics[scale=0.43]{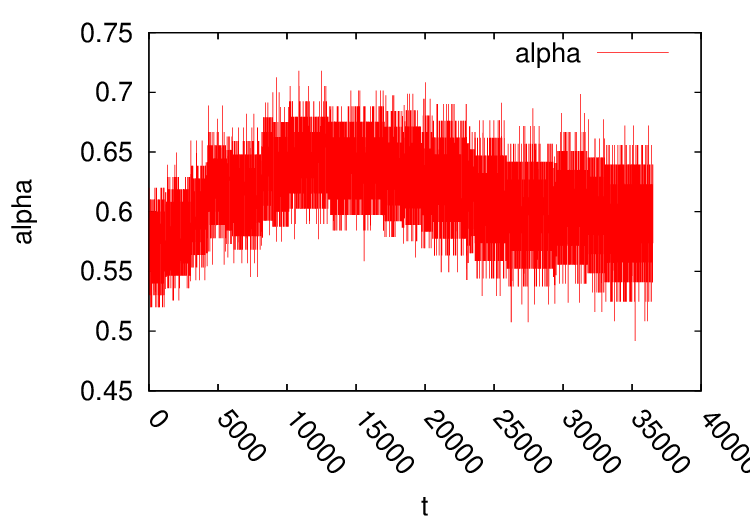}(c)
\includegraphics[scale=0.43]{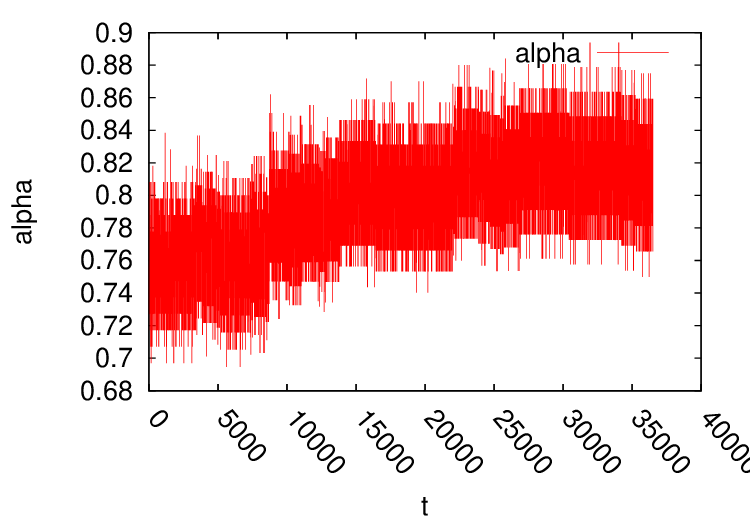}(d)
\caption{COLOR ONLINE. 
The temporal development of $\alpha(t)$; \textcolor{black}{(a) 
contrarians-dominant market, (b) contrarians-predominant-market, 
(c) trend-followers-predominant market, and (d) trend-followers-dominant 
market}.}
\label{fig:alpha}
\end{figure}

Fig. \ref{fig:markets} shows the market prices at (a) 
\textcolor{black}{$\alpha(0)=0.31$, (b) $\alpha(0)=0.48$, 
(c) $\alpha(0)=0.56$, and (d) $\alpha(0)=0.75$.
$\alpha(0)=0.31$ (contrarians-dominant market) represents a case where
contrarians are dominant in the market. $\alpha(0)=0.48$
(contrarians-predominant-market) corresponds to a
case where contrarians are still dominant but trend followers are more
in number than in the case of $\alpha(0)=0.31$. In the case of
$\alpha(0)=0.56$ 
(trend-followers-predominant market), trend followers are more than 
contrarians in the market. $\alpha(0)=0.75$ 
(trend-followers-dominant market) shows a case where 
trend followers are dominant in the market. We compare the four 
cases and examine dependency of $\alpha(t)$ on the market condition.
} 
\textcolor{black}{Tab. \ref{tab:markets} shows descriptive statistics of
  the market prices for the four cases.  As shown in Fig. \ref{fig:alpha}, we confirmed that $\alpha(t)$
  fluctuates with some variance and slightly varies in time due to
  a bankruptcy of banks. Namely, if a bank goes bankrupt at time $t$, then 
  $N(t+1) = N(t) - 1$. If it is further a trend follower, then
  $N_{tf}(t+1) = N_{tf}(t) -1$.}

\begin{figure}[!hbt]
\centering
\includegraphics[scale=0.43]{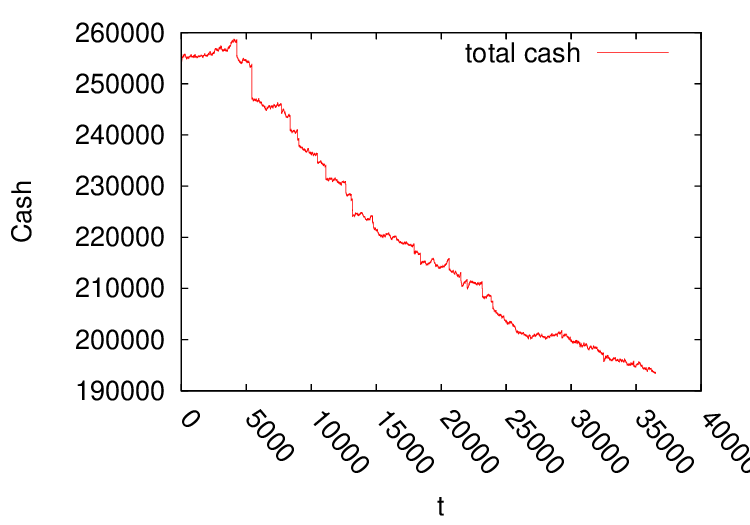}(a)
\includegraphics[scale=0.43]{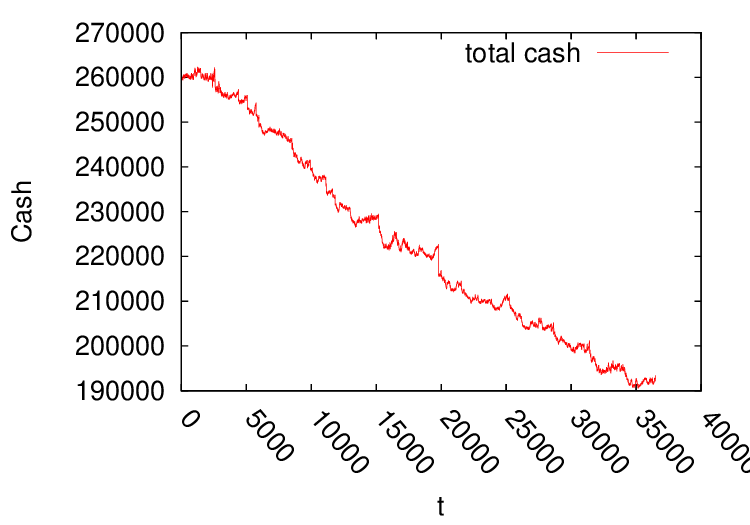}(b)
\includegraphics[scale=0.43]{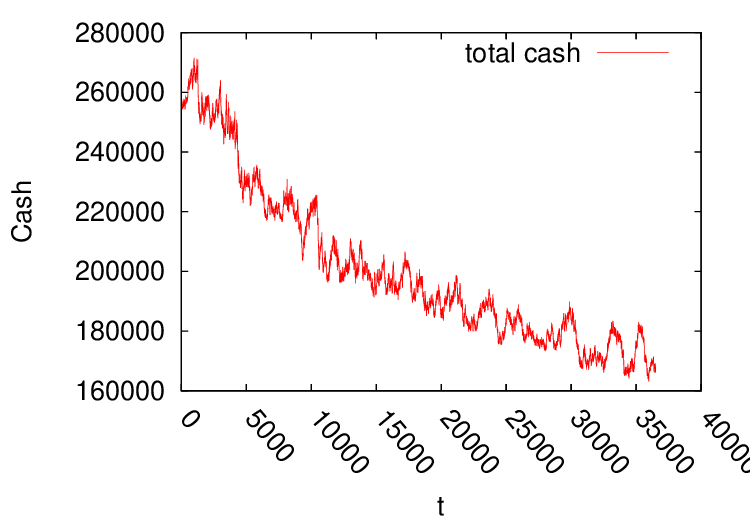}(c)
\includegraphics[scale=0.43]{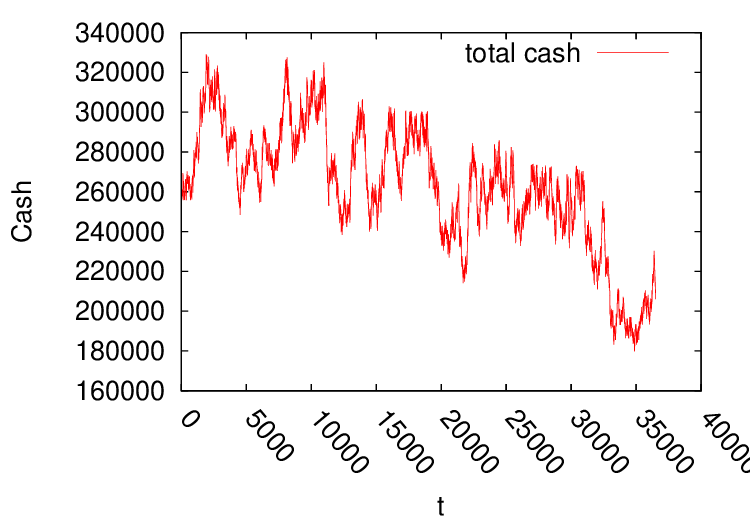}(d)
\caption{COLOR ONLINE. 
The total amount of cash; \textcolor{black}{(a) contrarians-dominant market, 
(b) contrarians-predominant-market, (c) trend-followers-predominant market, 
and (d) trend-followers-dominant market}.}
\label{fig:cashs}
\end{figure}

\begin{figure}[!hbt]
\centering
\includegraphics[scale=0.43]{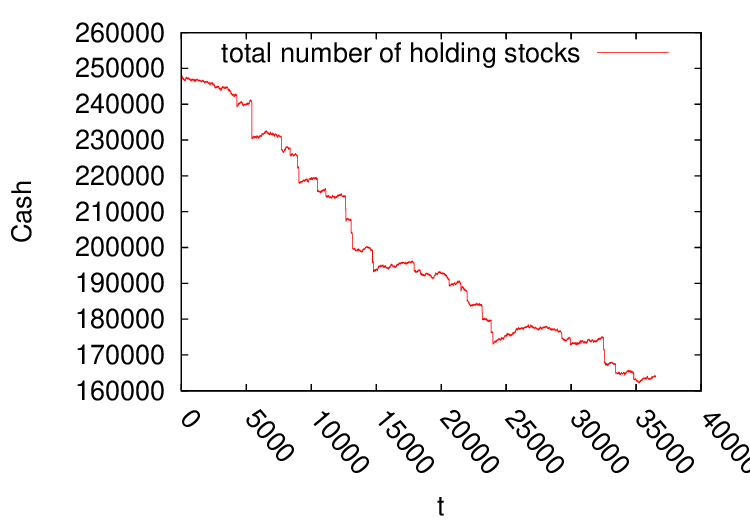}(a)
\includegraphics[scale=0.43]{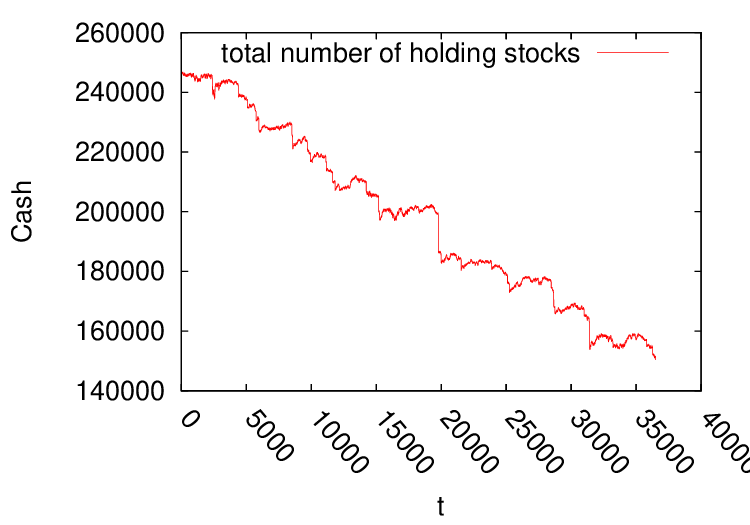}(b)
\includegraphics[scale=0.43]{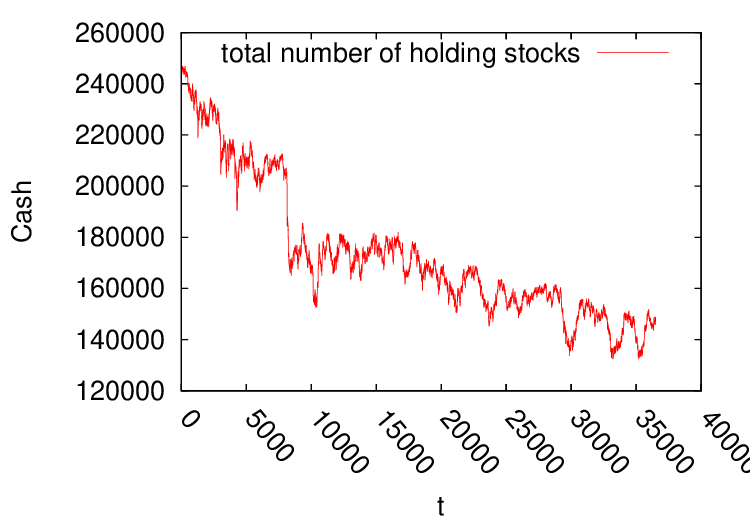}(c)
\includegraphics[scale=0.43]{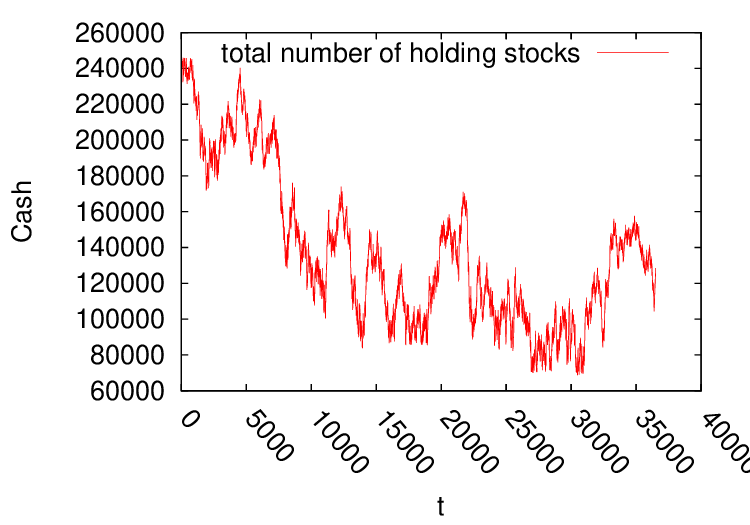}(d)
\caption{COLOR ONLINE. 
(a) The total amount of risk assets; \textcolor{black}{(a) contrarians-dominant 
market, (b) contrarians-predominant-market, (c) trend-followers-predominant 
market, and (d) trend-followers-dominant market}.} 
\label{fig:stocks}
\end{figure}

The duration of these plots corresponds to 100 years (36,500 days).
Fig. \ref{fig:cashs} shows the total amount of cash and
Fig. \ref{fig:stocks} shows the total amount of risky assets. \textcolor{black}{Tabs. \ref{tab:cashs} and \ref{tab:stocks} show descriptive statistics of the total amount of cash and risky assets.}

When the market price goes down, the total amount of risky assets
decreases and the total amount of losses increases. This situation can
be confirmed at \textcolor{black}{$\alpha(0)=0.75$}.

\begin{table}[!hbt]
\centering
\textcolor{black}{
\caption{Descriptive statistics of the market prices for four cases.}
\label{tab:markets}
\begin{tabular}{rrrrrr}
\hline
\multicolumn{1}{c}{$\alpha(0)$} & \multicolumn{1}{c}{Mean} & \multicolumn{1}{c}{Median} & \multicolumn{1}{c}{Var} & \multicolumn{1}{c}{Min.} & \multicolumn{1}{c}{Max.} \\
\hline
0.31 & 0.396906 & 0.388796 & 0.002148 & 0.288575 & 0.925797 \\
0.48 & 0.422041 & 0.419790 & 0.003362 & 0.276761 & 0.890119 \\
0.56 & 0.268880 & 0.260136 & 0.015824 & 0.025530 & 0.876253 \\
0.75 & 0.024344 & 0.001470 & 0.004728 & 0.000013 & 0.838115 \\
\hline
\end{tabular}
}
\end{table}

\begin{table}[!hbt]
\centering
\textcolor{black}{
\caption{Descriptive statistics of the total amount of cash for four cases.}
\label{tab:cashs}
\begin{tabular}{rrrrrr}
\hline
\multicolumn{1}{c}{$\alpha(0)$} & \multicolumn{1}{c}{Mean} & \multicolumn{1}{c}{Median} & \multicolumn{1}{c}{Var} & \multicolumn{1}{c}{Min.} & \multicolumn{1}{c}{Max.} \\
\hline
0.31 & 221025.65 & 216784.15 & 434426403.00 & 193298.04 & 258798.49 \\
0.48 & 222981.91 & 221224.41 & 471846434.54 & 190538.46 & 262366.22 \\
0.56 & 200522.00 & 193584.94 & 686616124.42 & 163129.25 & 271543.81 \\
0.75 & 262675.93 & 265049.78 & 959870096.96 & 179773.39 & 329123.67 \\
\hline
\end{tabular}
}
\end{table}

\begin{table}[!hbt]
\centering
\textcolor{black}{
\caption{Descriptive statistics of the total amount of risk assets for four cases.}
\label{tab:stocks}
\begin{tabular}{rrrrrr}
\hline
\multicolumn{1}{c}{$\alpha(0)$} & \multicolumn{1}{c}{Mean} & \multicolumn{1}{c}{Median} & \multicolumn{1}{c}{Var} & \multicolumn{1}{c}{Min.} & \multicolumn{1}{c}{Max.} \\
\hline
0.31 & 199288.73 & 193253.00 & 711326678.75 & 162122.00 & 253637.00 \\
0.48 & 198077.09 & 199944.00 & 827698672.91 & 150323.00 & 253288.00 \\
0.56 & 173392.31 & 167072.00 & 710520661.60 & 132277.00 & 253255.00 \\
0.75 & 137052.78 & 128987.00 & 1839165861.20 & 68731.00 & 253252.00 \\
\hline
\end{tabular}
}
\end{table}

If the ratio of the contraians to the total traders increases, 
then price movement seems to be mean-reverting. \textcolor{black}{As shown 
in Tab. \ref{tab:markets}, the variance of the contrarians-dominant-market 
at $\alpha(0)=0.31$ is the smallest in the four cases. Other cases 
($\alpha(0)=0.56$ and $0.75$) show higher volatilities and correspond 
procyclical market behavior between bubbles and crashes. As shown in 
Tabs. \ref{tab:cashs} and \ref{tab:stocks} it is shown that the total 
amounts of cash and risky assets at $\alpha(0)=0.56$ become less volatile 
than other cases. This implies that the total amounts of cash and risky 
assets in the banking system of the trend-followers-predominant market 
at $\alpha(0)=0.56$ change less drastically than other cases 
($\alpha(0)=0.75$). Namely, if trend followers are dominant in the market, 
then the procyclical behavior of the market becomes harmful for market 
participants and it may make banking systems more unstable.}

\begin{figure}[!hbt]
\centering
\includegraphics[scale=0.67]{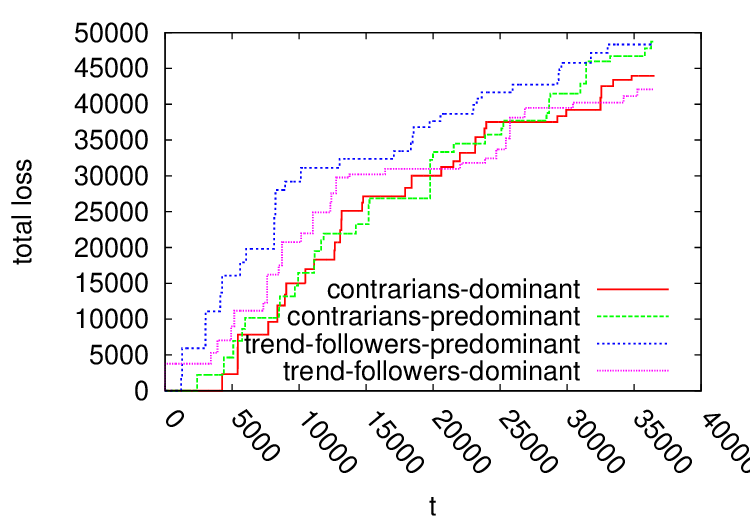}
\caption{COLOR ONLINE. The total amount of economic losses $H(t)$ \textcolor{black}{for the four cases.}}
\label{fig:loss}
\end{figure}

The cumulative losses $H(t)$ is shown in Fig. \ref{fig:loss}. 
\textcolor{black}{It is said that if market participants are
homogeneous, the total amount of losses is less than more heterogeneous
cases. In both the trend-followers-predominant and
contrarians-predominant markets, the number of bankruptcy is larger
than the two other cases such as the trend-followers-dominant and
contrarians-dominant markets. Contrarians may obtain profit from the
market when the market price is mean-reverting. Trend followers may
obtain profit from the market when the market price is procyclical.}

\begin{figure}[!hbt]
\includegraphics[scale=0.43]{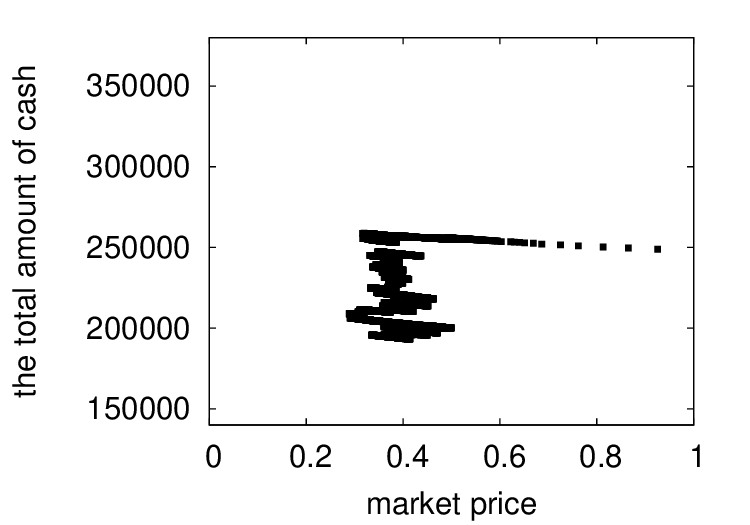}(a)
\includegraphics[scale=0.43]{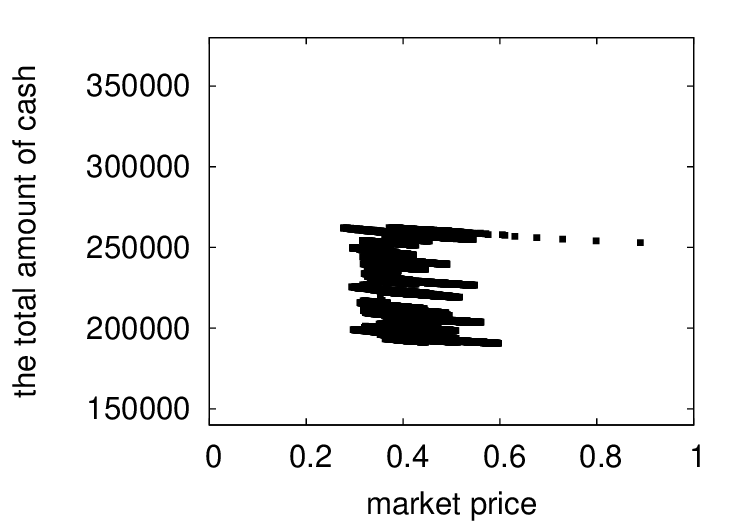}(b)
\includegraphics[scale=0.43]{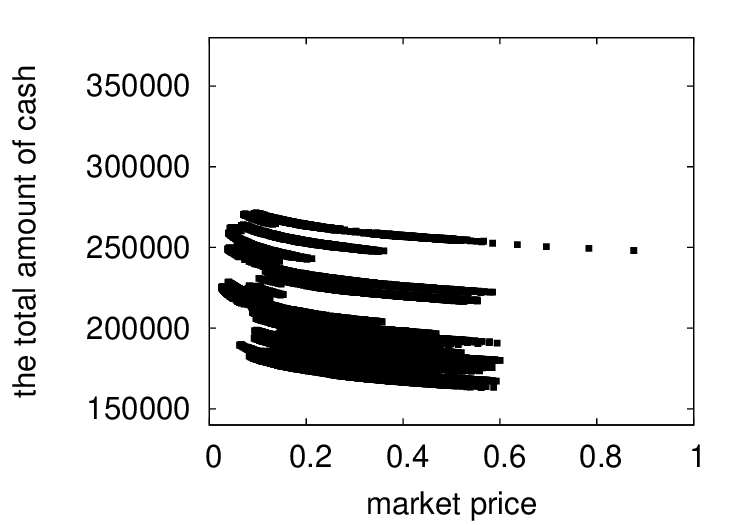}(c)
\includegraphics[scale=0.43]{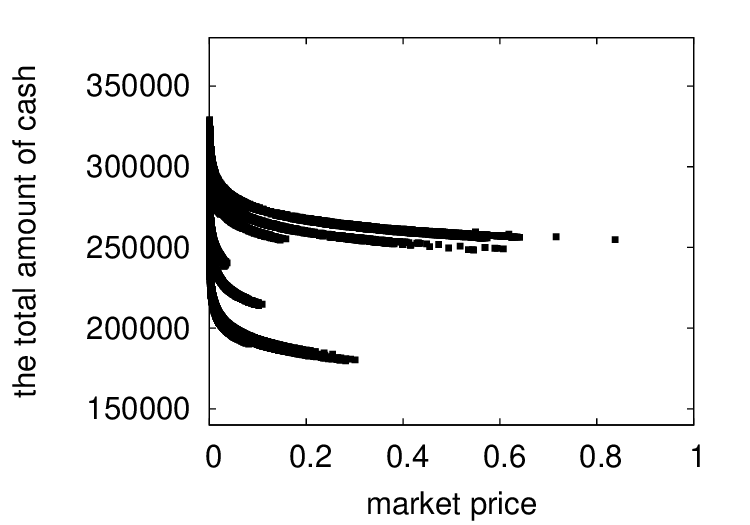}(d)
\caption{Scatter plots between the market prices and the total amount
  of cash; \textcolor{black}{(a) contrarians-dominant market, (b) contrarians-predominant-market, (c) trend-followers-predominant market, and (d) trend-followers-dominant market}.}
\label{fig:comovement1}
\end{figure}

\begin{figure}[!hbt]
\includegraphics[scale=0.43]{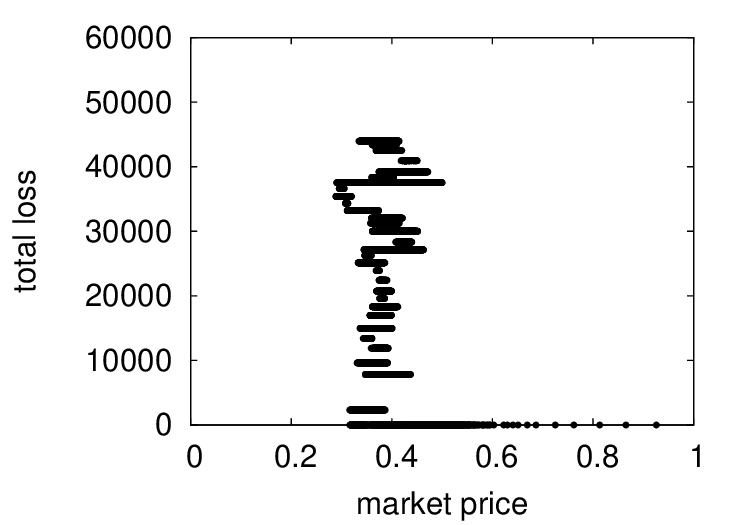}(a)
\includegraphics[scale=0.43]{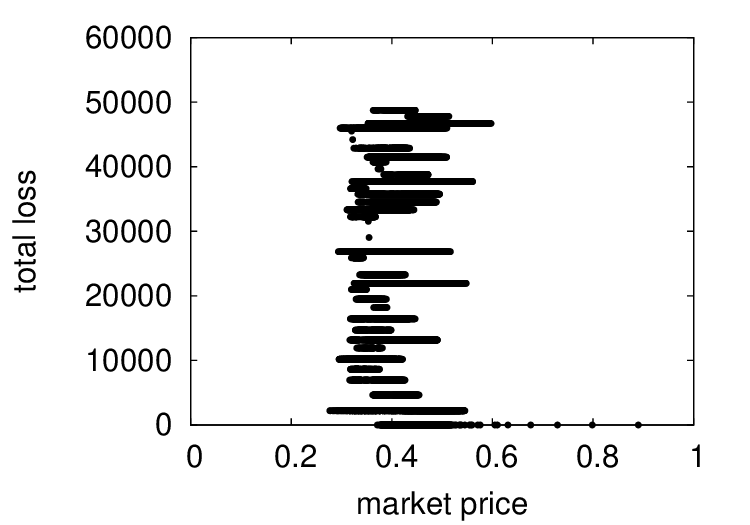}(b)
\includegraphics[scale=0.43]{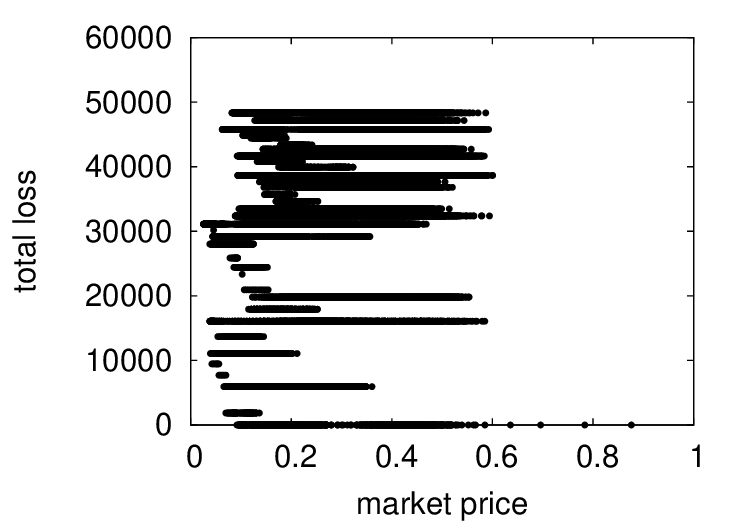}(c)
\includegraphics[scale=0.43]{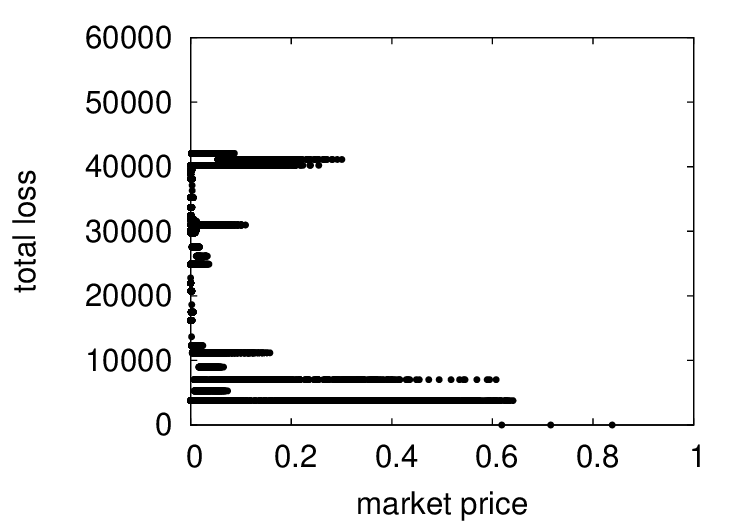}(d)
\caption{Scatter plots between the market prices and the total losses;
\textcolor{black}{(a) contrarians-dominant market, (b) contrarians-predominant-market, (c) trend-followers-predominant market, and (d) trend-followers-dominant market}.}
\label{fig:comovement2}
\end{figure}

\textcolor{black}{Figs. \ref{fig:comovement1} and \ref{fig:comovement2}
show scatter plots between the market prices and the total amount
of cash and those between the market prices and the total losses.
We found that mean-reverting price movements may be less harmful 
to banks than high volatile price movements.}

\begin{figure}[!hbt]
\centering
\includegraphics[scale=0.43]{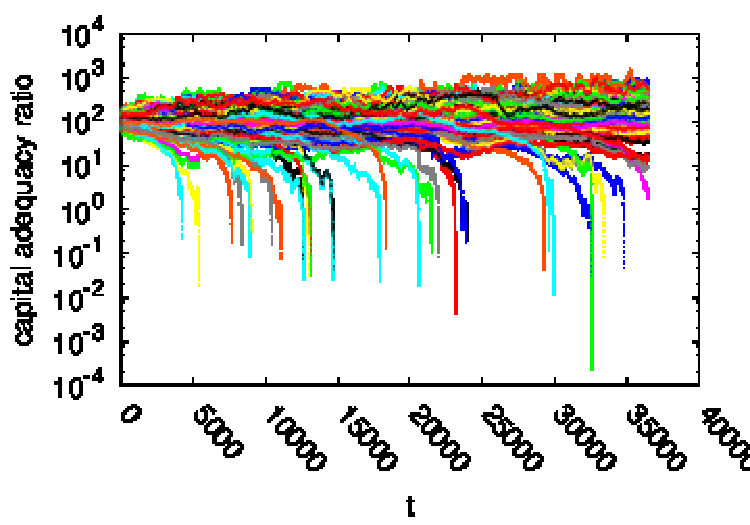}(a)
\includegraphics[scale=0.43]{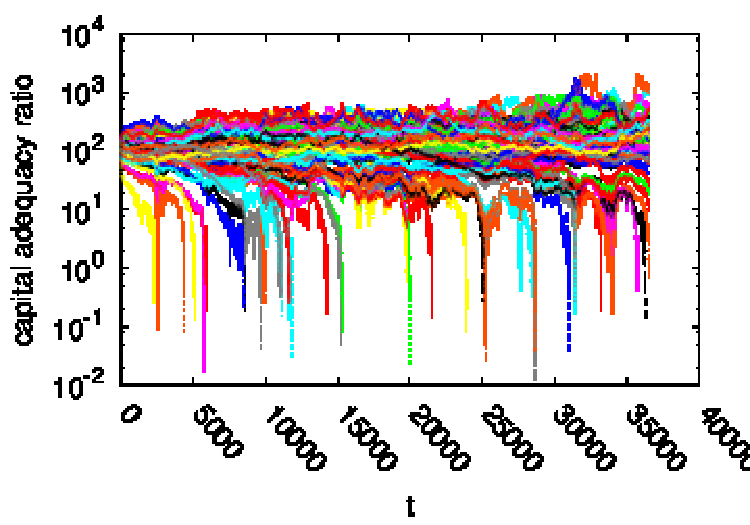}(b)
\includegraphics[scale=0.43]{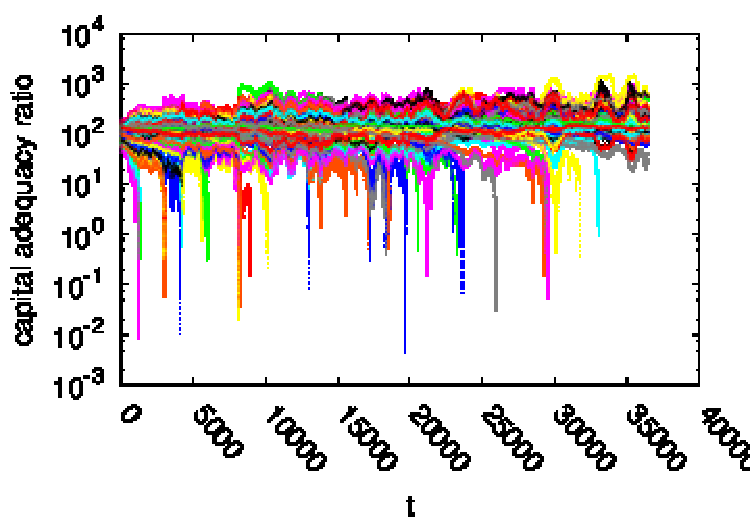}(c)
\includegraphics[scale=0.43]{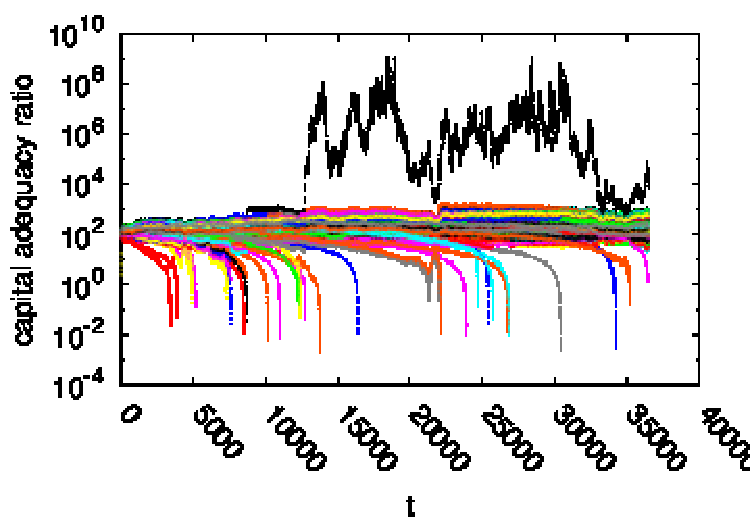}(d)
\caption{COLOR ONLINE. 
Capital adequacy ratio of every bank; \textcolor{black}{(a)
contrarians-dominant market, (b) contrarians-predominant-market, (c)
trend-followers-predominant market, and (d) trend-followers-dominant
market}.}
\label{fig:CAR-banks}
\end{figure}

Fig. \ref{fig:CAR-banks} shows time series capital adequacy ratio (CAR) at
\textcolor{black}{(a) $\alpha(0)=0.31$, (b) $\alpha(0)=0.48$, (c)
  $\alpha(0)=0.56$, and (d) $\alpha(0)=0.75$}. It is confirmed that
the banks which went bankrupt have small CAR. Before the banks went
bankrupt, value of CAR \textcolor{black}{dropped} steeply. Therefore, the default
probability should be a function of CAR, which is not homogeneous. In
general, a market where contrarians are dominant shows mean-reverting
price movements, and a market where trend followers are dominant makes the
market price volatile. However, \textcolor{black}{the} CAR of the
contrarians tends to decrease and \textcolor{black}{the} CAR of the
trend followers tends to increase.

\begin{figure}[!hbt]
\begin{center}
\includegraphics[scale=0.7]{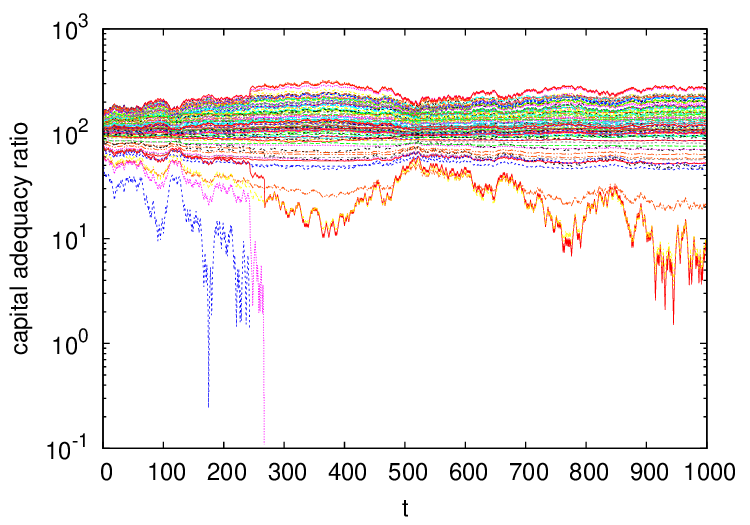}(a)
\includegraphics[scale=0.7]{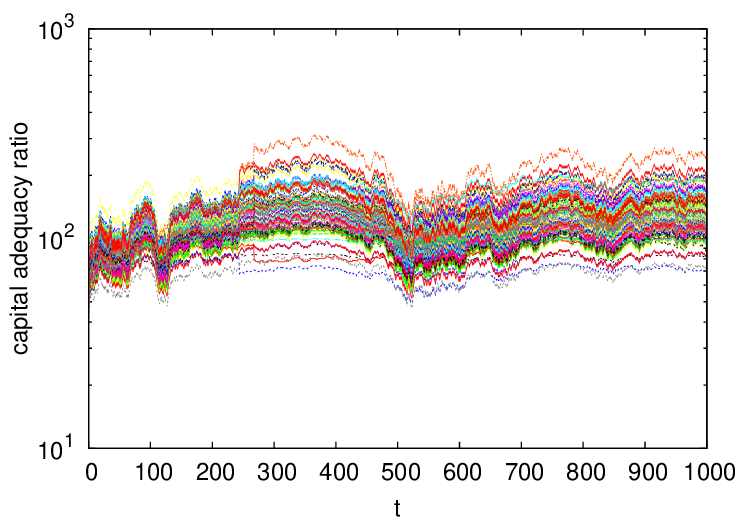}(b)
\caption{COLOR ONLINE. Temporal development of \textcolor{black}{(a)} CAR and \textcolor{black}{(b)} CEAR at \textcolor{black}{$\alpha(0)=0.79$}.}
\label{fig:CAR-CEAR}
\end{center}
\end{figure}

We compare sensitivity of \textcolor{black}{the} CAR and that of 
\textcolor{black}{the} CEAR for some cases.  
Figure \ref{fig:CAR-CEAR} shows temporal developments of CAR and CEAR for
100 banks (\textcolor{black}{$\alpha(0)=0.79$}). When banks go bankrupt,
their CAR decreases eventually, however, their CEAR does not
change. This implies that \textcolor{black}{the} CAR can be used as an
indicator to measure banks condition but \textcolor{black}{the} CEAR
might not be used as an indicator for such a purpose. The sensitivity
of CAR is extremely better than CEAR. We should recognize the
difference of characteristics between CAR and CEAR.

\section{Discussion}
\label{sec:discussion}

Our approach has several limitations. The first problem is related to
relationships between simple and complex agent-based
model. \textcolor{black}{Usually},
simple agent models tend to be too simple to apply actual risk
estimation. The main purpose of simple agent-based models is to
specify roles and components assumed in phenomena which we want to
draw. Through a modeling process, we eventually understand
\textcolor{black}{the} structure
of \textcolor{black}{these} problems and \textcolor{black}{we} identify
roles and components. In our case, we use 
simple agent-based model to identify fundamental relationships among
agents and roles of financial assets in balance sheets.  

Meanwhile, the main purpose of complex agent-based
\textcolor{black}{models} is to capture, reproduce and simulate
phenomena. To do so, we also consider 
how to calibrate model parameters. Generally speaking, it is not easy
to estimate all the parameters under a reliable procedure. We
sometimes face parameters' ambiguity to turn out similar
results. Namely, similar results can be created by different sets of
parameters. This problem is related to nonlinearity of parameters. 

It is known that a simple agent-based model tends to become a complex
agent-based model through a process of \textcolor{black}{improvement}. If we
improve our model, eventually the purpose of our model changes from
structure specification to risk estimation. In this case, we will also
suffer from calibration problems. 

The second limitation of our approach is parameter calibration. We do
not have sufficient knowledge on actual banking systems. In fact,
partial data of financial and banking systems can be
used\textcolor{black}{;} 
however, we do not have all the data to calibrate model parameters. Furthermore,
our agent-based model is too simple to apply risk assessment of
actual banking networks. Even though it is simple, we can understand
interplay between banks and to use it for developing and testing
indicators. For example, we can recognize that financial assets have a
potential to play a role of a common factor and cash adequacy ratio
and capital adequacy ratio can be used as measures to
estimate. Furthermore, if we can calculate their correlations in terms
of banks, we can quantify intensity of common factors.

The third limitation is related to \textcolor{black}{the}
expressions \textcolor{black}{of the model}. We used a matrix   
to describe relationships between
banks. However, \textcolor{black}{banks} sometimes appear and disappear
due to new launch, bankruptcy and M\&A. The matrix representation
cannot express such things.

However, our \textcolor{black}{aim} is to identify important components in the
bank balance sheet and \textcolor{black}{have deeper understanding}
among them. Although we model a simplified version of the banking and
financial systems, we established a  
useful benchmark reference model that \textcolor{black}{clarifies how the 
feedback loop between bank behaviors and how asset prices impact of
default risk}. 

According to BIS consolidated banking statistics in
2015Q1~\cite{BISConsolidated}, total assets for all bank nationalities
is 70,082.7 (billions of USD), the total amount of loans and deposits
is 65,919.9 (billions of USD) and the total amount of debt securities
is 7,862,0 (billions of USD) (see Table \ref{tab:BISConsolidated}). We
found that the total amount of assets is larger than liabilities for
debt securities. Namely, we may justify our hypothesis that the
exposure to financial markets is larger than amounts \textcolor{black}{lent} and
borrowed in an interbank network.
 
\begin{table}[!hbt]
\centering
\caption{BIS consolidated banking statistics in 2015Q1. The amount is totaled
over all CSB-reporting banks. Amounts outstanding, in billions of US dollars.}
\label{tab:BISConsolidated}
\begin{tabular}{lr}
\hline
items & amount \\
\hline 
Foreign claims (Immediate counterparty) & 27,077.7 \\
Foreign claims (Ultimate risk) & 24,231.7 \\
Domestic claims (Immediate counterparty) & 46,882.2 \\
Domestic claims (Ultimate risk) & 46,627.4 \\
Total assets & 70,082.7 \\
Liabilities (Total) & 65,919.9 \\
Liabilities (Of which Loans and deposits) & 45,095.7 \\
Liabilities (Of which Debt securities) & 7,862.0 \\
Liabilities (Of which Derivatives) & 6,336.0 \\
Total equity & 4,809.1 \\
\hline
\end{tabular}
\end{table}

\section{Conclusion and future work}
\label{sec:conclusion}
We emphasized the fact that in our \textcolor{black}{suggested} model we
were able to 
capture the relation between banks behavior and asset prices. We
described the positive feedback-loop between banks default probability
and asset price dynamics. The results showed that the procyclical
banks' behaviors (i.e., feedback loop between default probability and asset
prices) can explain the realization of asset price bubbles and their
burst. The characteristic of the interbank market plays in this
context a minor role if exposures in financial markets
are larger than capital in an interbank network. From this view, the
interbank market simply allows us to condense the individual
default probabilities into the systemic default probability. 
The capital adequacy ratio (Leverage ratio) is a useful indicator to
monitor the default probability.

As \textcolor{black}{for the} future, we need to check our hypothesis
against detailed data 
and estimate probability of procyclicality. \textcolor{black}{The model 
parameters depend on the macroscopic behavior of financial markets and bank default
probability obviously. However, the parameters strongly depend also on
market conditions and banks risk appetite. Unfortunately, we do not have
enough information about bank risk appetite in general. We need to
infer macro economic conditions, bank balance sheets, and market
sentiment. Our paper paves the way to discuss this important issue in
further studies. Therefore, parameter estimations from available information
about actual banks' behavior are crucial future tasks.}


\section*{Conflict of interests}
The authors declare no conflict of interest associated with
this manuscript. Views expressed are those of authors and do not
necessarily reflect those of the Bank (Bundesbank, Bank of Japan). 

\section*{Acknowledgment}
This work was financially supported by a Grants-in-Aid for Scientific
Research (KAKENHI) (B) (\#26282089). 



%

\end{document}